\documentclass[prb,amssymb,amsmath,twocolumn,showpacs]{revtex4-1}
\usepackage{bm}
\usepackage{graphicx}
\usepackage{color}
\usepackage[normalem]{ulem}
\usepackage{hyperref}

\begin{document}

\title{Topologically-protected entanglement of electron pair cyclotron motions}
%\title{\textcolor{red}{Hidden topological properties of entangled two-body wave functions \\ in two-dimensional electron gases under high magnetic fields}}

\author{T. Champel}
\affiliation{Universit\'e Grenoble Alpes/CNRS, Laboratoire de Physique et Mod\'elisation des Milieux Condens\'es, B.P. 166, 38042 Grenoble Cedex 9, France}

\author{D. Hernang\'{o}mez-P\'{e}rez}
\affiliation{Universit\'e Grenoble Alpes/CNRS, Laboratoire de Physique et Mod\'elisation des Milieux Condens\'es, B.P. 166, 38042 Grenoble Cedex 9, France}

\author{S. Florens}
\affiliation{Institut N\'{e}el, CNRS and Universit\'{e} Grenoble Alpes, B.P. 166, 38042 Grenoble Cedex 9, France}

\begin{abstract}

Considering two-dimensional  electron gases under a perpendicular magnetic field, we pinpoint a specific kind of long-range bipartite entanglement of the electronic motions. This entanglement is achieved through the introduction of bicomplex spinorial eigenfunctions admitting a polar decomposition in terms of a real modulus and three real phases.
Within this bicomplex geometry the cyclotron motions of two electrons are intrinsically tied, so that the highlighted eigenstates of the kinetic energy operator actually describe the free motion of a genuine electron pair. Most remarkably, these states embody phase singularities in the four-dimensional  (4D) space, with singular points corresponding to the simultaneous undetermination of the three phases.
% In addition to the locations of these topological defects giving rise to a continous 4D quantum number, there are two integral quantum numbers, which play a role analogous to the winding numbers for complex vortices. These numbers somehow describe here two rotational modes in opposite directions,  where the mode with the same direction of rotation as the classical cyclotron motion is associated with the pair Landau level index. Finally, a fourth quantum number is given by a parity index related to the spinorial structure of the states, which guarantees the real positivity of the density probability in the complex Hilbert space spanned by the basis of the 4D topological defects.
Because the entanglement between the two electrons forming a pair, as well as the winding and parity quantum numbers characterizing the 4D phase singularity, are topological in nature, we expect them to manifest some robustness in the presence of a smooth disorder potential and an electron-electron interaction potential. The relevance of this effective approach in terms of 4D vortices of electron pairs is discussed in the context of the fractional quantum Hall effect.

\end{abstract}

\date \today


\maketitle

\section{Introduction}

One of the most important effects  of a strong perpendicular magnetic field on  electrons moving in a two-dimensional plane (2D) is to bend their trajectories into circular cyclotron orbits as a result of the Lorentz force. A byproduct of this classical cyclotron motion  is the generation of an important energy degeneracy with respect to the location of the center of the orbit  in the plane. In the quantum realm, the circular motion implies that the electronic kinetic energy becomes quantized into discrete and macroscopically degenerate Landau levels $E_{n_1}=(n_1+1/2)\hbar \omega_c$ where $\omega_c$ is the cyclotron pulsation and $n_1$ a positive integer, as shown by Landau in the early days of quantum mechanics \cite{Prange}. This quantization has especially dramatic consequences for transport properties of electronic systems in reduced dimensionality, the most famous and spectacular manifestations being indubitably the integral and fractional quantum Hall effects \cite{Klitzing1980,Tsui1982}.

To understand microscopically these effects \cite{Prange}, one must inherently deal with the large degeneracy of the Landau levels, which is lifted by the potentials due to the impurities and the interactions between electrons. This physical problem proves to be highly complex at high magnetic fields owing to its nonperturbative nature after projection onto a given Landau level. Concomitantly, the enormous degeneracy of the Landau levels implies a great freedom in the choice of a basis of eigenstates of the kinetic energy operator to study the process of degeneracy lifting.  
It has been realized during the past few years \cite{Raikh1995,Champel2007,Champel2008,Champel2009,Champel2010} that for the single electron problem (in which the interaction between electrons is treated at the mean field level) one peculiar set of vortex eigenstates turns out to be well suited  to capture the effects of an arbitrary non-uniform potential energy by virtue of its topological characteristics. Indeed, within the vortex state basis representation, the degeneracy quantum number  corresponds to the positions in the plane of 2D phase singularities for the single electron wave function, while the Landau level index $n_1$ acquires the meaning of a (positive) circulation around the same singularities. %These two quantum numbers describing the electronic quantum motion in the plane are thus not entirely independent in so far as they are both required to fully characterize the same topological defect. 

The quantized circulation  embodies the ability of the defect to be preserved at high magnetic fields under an arbitrary continuous potential energy perturbation, whose main effect is to confer to the vortex a slow motion in an effective one-body potential landscape. The incompressible nature of this vortex flow\cite{Champel2007} is then responsible for the quantization of the Hall conductance. 
%sets in a quasi-ballistic transport regime, with the localization of all the bulk states except at the center of a Landau level where a propagation throughout the system can take place only via percolation, as recently confirmed by scanning tunneling spectroscopy experiments in the integer quantum Hall regime.
At the theoretical level, the phase space quantization with respect to the position of the phase singularity  (which can be identified in the semiclassical limit as the orbit guiding center) is achieved through the coherent states algebra. Consequently, the degeneracy lifting by a non-uniform potential energy 
%giving rise to a smooth deformation of the vortex states 
is accounted for in a differential way \cite{Champel2009,Champel2010}, which allows the incompressible vortex flow to adjust locally. Because the quantum motion of the vortex is associated with relatively small energy scales at high magnetic fields, one can then typically devise semiclassical-type approximations to describe quantitatively its propagation.

In this paper, we generalize at the two-electron level this association of the Landau quantization with the existence of a singular incompressible flow within an hydrodynamical picture of quantum mechanics \cite{Madelung1926,Taka1983}, with the aim to describe afterwards the degeneracy lifting by an interaction two-body potential. In this problem, where the number of degrees of freedom is now doubled, %(i.e., the quantum states are necessarily distinguished by  four quantum numbers)
 the non-interacting two-electron system with the energy quantization $(n_1+n_2+1)\hbar \omega_c$ obviously presents a higher level degeneracy, hinting at an even greater freedom in the choice of a basis of eigenstates than in the single electron case.
This means again that the Landau quantization can be interpreted in many different ways, depending on the chosen decomposition for the global motion of the two electrons into elementary motions.

 Anticipating a high-magnetic field projection onto two-electron Landau levels, one may  also wonder whether it is possible to regard the sum $n_1+n_2$ of the individual Landau level indices as a single quantum number which has the meaning of a circulation in some peculiar representation of the quantum eigenstates. Such a speculation immediately entails searching for entangled eigenstates for which the two electrons have lost their individuality and can not be seen as separable entities. The corresponding eigenstates would then describe a genuine pair of electrons. 
The main result of this paper is to pinpoint such a basis of pair eigenstates  describing a four-dimensional (4D) singular hydrodynamic incompressible flow. We show that, unlike the single-electron case, the three remaining degeneracy quantum numbers describing the point-like defect in the pair motion space have a mixed continuous and discrete character. Because of their topological origin, the discrete degeneracy quantum numbers should exhibit some robustness at the same time that they are revealed when switching on an arbitrary two-body interaction. In other terms,  the 4D vortices of electron pairs are expected  to form stable quasiparticles at high magnetic fields provided that their constitution will be protected by large enough gaps of the resulting effective two-body interaction.

This paper is organized as follows. In Sec. \ref{2D}, we briefly review the main properties of the complex vortex basis for the single electron problem. This section will also serve as an introduction to the construction method of the bicomplex pair vortex eigenstates, which bear many similarities to their 2D analogues.
 While the underlying bicomplex algebra and the topological properties of the  pair eigenstates are exposed in Sec. \ref{4D}, the completeness relation obeyed by these states and the structure of the associated Hilbert space are addressed in Sec. \ref{basis}. In Sec. \ref{pers} we eventually discuss a promising application of the formalism, by arguing for the relevance of considering 4D vortices of electron pairs as building blocks of an effective theory for the fractional quantum Hall effect.
 Our work is summarized in the conclusion (Sec. \ref{conc}). Finally, some of the technical details are collected in the Appendixes.

\section{One-body vortex eigenstates \label{2D}}

We first consider a single electron of effective mass $m^{\ast}$ and of charge $e=-|e|$ freely moving in a two-dimensional plane $(x_1,y_1)$ under a perpendicular magnetic field ${\bf B}=B \hat{{\bf z}}$. The Hamiltonian then simply consists of the kinetic energy operator
\begin{eqnarray}
\hat{H}_{0}({\bf r}_1)= \frac{1}{2 m^{\ast}} \left(-i \hbar {\bm \nabla}_{{\bf r}_1}+\frac{|e|}{c} {\bf A}({\bf r}_1) \right)^2, \label{Ham1el}
\end{eqnarray}
where ${\bf A}({\bf r}_1)$ is the vector potential, defined up to a gauge factor by the equation
\begin{eqnarray}
{\bm \nabla}_{{\bf r}_1} \times {\bf A}({\bf r}_1)={\bf B}.
\end{eqnarray}
There exist many different ways to derive the  corresponding well-known quantized Landau energy spectrum, which reads $E_{n_1}=(n_1+1/2)\hbar \omega_c$ with the cyclotron pulsation defined as $\omega_c=|e|B/(m^{\ast}c)$. As a result, this Landau level quantization may be interpreted from diverse viewpoints (as a consequence of the square integrability condition, of the quantization of angular momentum, etc. \dots).
This is partly due to the macroscopically large degeneracy of the energy spectrum providing a great freedom in solving the stationary Schr\"{o}dinger's equation $\hat{H}_{0} \Psi =E \Psi$, which requires defining a second relevant (degeneracy) quantum number besides the integer $n_1$. 

However, it has been shown \cite{Champel2007} that one peculiar set of electronic eigenstates $\Psi_{n_1,{\bf R}_1}({\bf r}_1)=\langle {\bf r}_1  |n_1,{\bf R}_1  \rangle $ should be fundamentally preferred \cite{Champel2009,Champel2010} in so far as it embodies a hydrodynamic incompressible vortex flow  in which the Landau level quantization follows from a purely topological condition. This appears as a direct consequence of the presence of a magnetic field, which unavoidably leads to the existence of a non-trivial phase $\varphi({\bf r}_1)$ for the complex wave function $\Psi(\mathbf{r}_1)= |\Psi(\mathbf{r}_1)| \exp [i \varphi(\mathbf{r}_1)]$ independently of the chosen gauge for the vector potential. Within a hydrodynamic  picture of quantum mechanics \cite{Madelung1926,Taka1983},    $m^{\ast}|\Psi(\mathbf{r}_1)|^2$ plays the role of a mass density and the (gauge-independent) quantity
\begin{eqnarray}
{\bf v}=\frac{\hbar }{m^{\ast}} \left( {\bm \nabla}_{{\bf r}_1} \varphi(\mathbf{r}_1)+\frac{2 \pi}{\Phi_0} {\bf A} (\mathbf{r}_1)\right)
\end{eqnarray}
corresponds to a  flow velocity (here $\Phi_0=h c/|e|$ indicates the magnetic flux quantum). One may then envision an incompressible flow  ${\bm \nabla}_{{\bf r}_1} \cdot {\bf v}=0$ displaying phase singularities, \textit{i.e.}, such that ${\bm \nabla}_{{\bf r}_1} \times {\bm \nabla}_{{\bf r}_1} \varphi({\bf r}_1)=2 \pi n_1 \, \delta \left({\bf r}_1-{\bf R}_1 \right) \hat{{\bf z}}$. We thus see that both quantum numbers $n_1$ and ${\bf R}_1$ share somehow a common origin since they both stem from the production of the same topological defect.

These  vortex wave functions, eigenstates of the (one-body) kinetic energy $\hat{H}_0({\bf r}_1)$, are explicitly given in the symmetrical gauge ${\bf A}({\bf r}_1)={\bf B} \times {\bf r}_1/2$ by the expression \cite{Champel2007,Malkin1969}

\begin{eqnarray}
\langle {\bf r}_1  |n_1,{\bf R}_1  \rangle 
%&= &\frac{1}{l_B \sqrt{2 \pi n_1!}} \left(\frac{z_1-Z_1}{\sqrt{2} l_B} \right)^{n_1} \, e^{-\frac{|z_1-Z_1|^2}{4 l_B^2}} \, e^{ \frac{i}{2 l_B^2} \left[ x_1Y_1 -y_1X_1 \right]} \nonumber \\&=& 
=\frac{l_B^{-1}}{\sqrt{2 \pi n_1!}} \left(\frac{z_1-Z_1}{\sqrt{2} l_B} \right)^{n_1} \, e^{-\frac{|z_1|^2+|Z_1|^2-2Z_1 z_1^{\ast}}{4 l_B^2}}, \hspace*{0.5cm} \label{vor}
\end{eqnarray}
where $l_B=\sqrt{\hbar c/(|e|B})$ is the magnetic length.
Here $z_1=x_1+iy_1$ is a complex number such that $(x_1,y_1)={\bf r}_1$ defines the electronic position in the plane. Similarly, ${\bf R}_1=(X_1,Y_1)$ is the vortex position in the 2D plane.  Importantly, the complex coordinate $Z_1=X_1+iY_1$ characterizes the location of the zeros of the wave function, which definitely correspond to phase singularities when $n_1 \geq 1$. Note that the Landau level index $n_1$ also characterizes the {\em positive} circulation around the vortex, which can be interpreted semi-classically as the chiral circling motion of the electrons with an axis pointing towards the field direction. The states \eqref{vor} should not be confused with the states usually considered in the context of the fractional quantum Hall effect for a projection on the lowest Landau level, which correspond to vortex solutions with {\em negative} circulation and all phase singularities located at the position ${\bf R}_1={\bf 0}$. Within the present convention of a magnetic field pointing in the $+ \hat{\bf z}$ direction, the lowest Landau level wave functions 
 would indeed exhibit an antiholomorphic character, since they only depend on the electronic variable $z_1^{\ast}=x_1-i y _1$ (if the global Gaussian factor $e^{-|z_1|^2/4l_{B}^2}$ is disregarded). 
The set of states \eqref{vor} spans the lowest Landau level eigenspace instead by considering an arbitrary position ${\bf R}_1$ in the plane and fixing the positive circulation $n_1=0$.

In fact, the vortex states \eqref{vor} obey the coherent states algebra \cite{Glauber1963,Zhang1990} with respect to the (doubly) continuous quantum number ${\bf R}_1$. Hence, a distinguishing feature is  that they present a non-orthogonal overlap 
\begin{eqnarray}
\langle n_1, {\bf R}_1 | n'_1, {\bf R}'_1 \rangle  = \delta_{n_1,n'_1} \langle {\bf R}_1 |{\bf R}'_1 \rangle , \label{overlap}
\end{eqnarray}
where
\begin{eqnarray}
 \langle {\bf R}_1 |{\bf R}'_1 \rangle =  e^{-\frac{|Z_1|^2+|Z'_1|^2-2 Z_1^{\ast} Z'_1}{4 l_B^2}}.
\end{eqnarray}
It can be easily shown \cite{Champel2007} that they nevertheless form a basis for the electronic quantum states, spanning the whole Hilbert space, with the completeness relation
\begin{eqnarray}
\int \frac{d^2 {\bf R}_1}{2 \pi l_B^2} \sum_{n_1=0}^{+ \infty} |n_1,{\bf R}_1 \rangle \langle n_1,{\bf R}_1 |=1\!\!1. \label{comp}
\end{eqnarray}
As it can be directly read from this relation, the degeneracy of the Landau levels is $(2 \pi l_B^2)^{-1}$ per unit area. 
In fact, the nonorthogonality with respect to the quantum number ${\bf R}_1$ in Eq. \eqref{overlap} reflects some freedom resulting from the overcomplete continuous character of the vortex basis, along with
the quantum uncertainty in the simultaneous determination of the vortex coordinates $X_1$ and $Y_1$, which form a pair of conjugate variables for non-zero magnetic field.
 % Actually, only the area of the quantized region in the $(X_1,Y_1)$ phase space is really fixed, whereas its definite shape will adjust locally according to the local configuration of the potential energy perturbation.  This important property can be seen in a way as the expression of stability and robustness of the vortex representation against any infinitesimal perturbations.
Obviously, the vortex positions ${\bf R}_1$ reduce to the classical guiding center coordinates in the limit where the magnetic length $l_B$ vanishes. The vortex representation $|n_1,{\bf R}_1 \rangle$  thus achieves, in a fully quantum mechanical language, the classical decomposition of the electronic motion into a cyclotronic rotation plus a guiding-center drift.

Note that the states \eqref{vor} have already been introduced at several occasions\cite{Girvin,Kivelson,Jain1987,Jain1988} in the context of the quantum Hall effect, essentially within a second-quantization language by making explicit reference to the coherent states definition. Within the hydrodynamic picture of quantum mechanics the two quantum numbers labeling the vortex eigenstates, the Landau level index and the guiding-center coordinates, proved to be not totally unrelated since they are both required to entirely characterize the topological defect. Thus, the coherent states character displayed by the vortex position turns out to be definitely a by-product. %The presence of a continuous (overcomplete) quantum number into the quantum description could also be viewed as some consequence of topology, where the concept of continuity forms a cornerstone.

The fact that the Landau level quantization process can be associated with the formation of a topological defect is a very important notion, which confers some rigidity properties to the vortex state representation. From purely topological grounds\cite{Mermin}, it is clear that the discrete vorticity quantum number cannot be changed easily. This principle translates into energetical terms by the presence of a gap (corresponding in the present case  to the Landau gap $\hbar \omega_c$) in the energy spectrum, which must be overcome in order to change the vorticity. 
This gap protection vindicates the Landau level projection, which turns out to be a good (perturbative) approximation to describe the electronic motion in an arbitrary smooth potential energy at sufficiently high magnetic fields.

Less obviously, the vortex states representation \eqref{vor} displays somehow an additional fundamental form of stability provided by the continuous character of the defect position ${\bf R}_1$. Indeed, the present overcomplete phase space formulation allows one to represent any state or operator in a diagonal form\cite{Sudarshan}, and, as a result, explicitly generates within the guiding center quantum dynamics a hierarchy of local energy scales\cite{Champel2009,Champel2010}  ordered by powers of the magnetic length and successive spatial derivatives of the potential energy. For a smooth potential energy, the vortex states are associated with the highest energy in this hierarchy and thus appear to be the most robust states, i.e., the most predictable ones in an experiment. The energy hierarchy then arranges the relevant superpositions of the vortex states by their degree of  nonlocality. It shows 
 that the passage from a purely local physics (characteristic of classical physics) to a highly nonlocal quantum physics  for the (slow) guiding center degree of freedom takes place gradually when one takes into account the presence of a decreasing low-energy cutoff into the physical description. This allows one to devise controlled semi-classical (nonperturbative) approximation schemes valid at {\em finite temperatures} for physical observables such as the thermal local density of states\cite{Champel2009,Champel2010}.  Hence, the global quantum mechanical motion of the electron in a high magnetic field and in the presence of a smooth potential energy can be effectively viewed as a moving vortex.

The possibility to write down an effective equation of motion for the vortex is based on the holomorphic character of the states \eqref{vor} with respect to the vortex positions. Indeed, the projection of the electron dynamics onto a given Landau level (not necessarily the lowest one) relies on the property that the states \eqref{vor} are analytical in the complex variable $Z_1$ for any Landau level (if we disregard the global Gaussian factor). Note that this property of analyticity, which is independent of the Landau level index, does not hold for the electronic variables ${\bf r}_1$, since the wave functions generically depend  both on the variables $z_1$ and $z_1^{\ast}$ (again once the Gaussian factor is removed). Because all Landau levels are treated on an equal footing, the vortex representation appears very convenient \cite{Champel2009,Champel2010} at the technical level to perform the Landau level projection.

The pioneer works\cite{Girvin,Kivelson,Jain1987,Jain1988} made use of the  states \eqref{vor} to study the lowest Landau level physics within a path-integral formalism, which seemed to suffer from technical difficulties that were not elucidated. In contrast, a formulation in terms of phase-space Green's functions\cite{Champel2008,Champel2009,Champel2010} is not tainted with the peculiar mathematical ambiguities often encountered with the path-integral technique. In the study of central interaction potentials, it seems appealing within the zero Landau level to introduce  vortex eigenstates with negative circulations instead of coherent states. However, this alternative for $n_1=0$ would mean giving up the continuous guiding center degree of freedom, which is albeit physically relevant at high magnetic fields. We show in the following  that there is more room at the two-electron level by building up vortex-like defects also within the lowest Landau level, which still exhibit some continuous character essential to describe non-perturbatively the physical effect of a smooth disordered potential on the electronic motion.

\section{Pair vortex eigenstates \label{4D}}

Let us consider now the two-electron kinetic energy operator consisting of the sum of the single-electron  free Hamiltonians 
\begin{eqnarray}
\hat{H}_{0}^{2e}({\bf r})=\hat{H}_{0}({\bf r}_1)+\hat{H}_{0}({\bf r}_2). \label{H2e}
\end{eqnarray}
Here ${\bf r}_1=(x_1,y_1)$ and ${\bf r}_2=(x_2,y_2)$ refer to the positions of the two electrons in the plane, while 
${\bf r}=({\bf r}_1,{\bf r}_2)$ forms the four-dimensional (4D) collection of these positions. Obviously, the corresponding two-electron energy levels take the form $E_{n_1}+E_{n_2}=(n_1+n_2 +1)\hbar \omega_c$ where $n_1$ and $n_2$ designate the individual Landau level indices. 

We are interested in solutions, $\Psi$, of the Schr\"{o}dinger's equation $\hat{H}_{0}^{2e}({\bf r}) \Psi({\bf r})=E \, \Psi({\bf r})$ which describe vortex-like defects. The product states $\Psi_{n_1,{\bf R}_1}({\bf r}_1) \Psi_{n_2,{\bf R}_2}({\bf r}_2)$ built from the one-electron vortex wave functions of each electron and introduced in the previous section are obviously eigenstates of $\hat{H}_{0}^{2e}({\bf r})$, but they embody two independent 2D vortices located at ${\bf r}_1={\bf R}_1$ and ${\bf r}_{2}={\bf R}_2$ with positive circulations $n_1$ and $n_2$. We are rather looking instead for eigenstates which can be viewed as 4D topological defects of the motion space for the two electrons, with the main goal of representing the integer $n_1+n_2$ as a single quantum number. We can guess that for this purpose the cyclotron motions of the two electrons have necessarily to be correlated with the achievement of a kind of closed trajectories in the 4D space. This also means that the sought eigenstates intrinsically describe the free motion of a pair of electrons rather than the motion of two (independent) free electrons. 

In the definition of the 2D complex vortex wave functions, the mathematical concepts of differentiability and analyticity play a major role. 
These important properties can be integrated in a 4D framework only if the space of solutions to the Schr\"{o}dinger's equation is extended to allow bicomplex-valued wave functions. Before presenting these 4D vortex solutions, we need to make a short presentation of the bicomplex algebra \cite{Davenport}, which is studied in detail, for instance, in a recent book by Catoni {\em et al.}\cite{Catoni2008}. 
 A bicomplex number $q$ can be originally viewed as a commutative extension of the complex numbers $z$ to the 4D space. It may be expressed as 
\begin{eqnarray}
q=z_1+jz_2=x_1+i y_1+j(x_2+iy_2),
\end{eqnarray}
  where $j$ is a hyperbolic unit, \textit{i.e.}, $j^2=1$. Here $i$ is the usual imaginary unit ($i^2=-1$) and commutes with $j$, hence we have $(ij)^2=(ji)^2=-1$. 
Somehow, the number $q$ can be seen as consisting of two copies of the complex plane correlated in a ``hyperbolic way''. 
In contrast to the usual Euclidean quaternions (which correspond to
a different, non-commutative, algebra on $\mathbb{R}^4$), the bicomplex numbers form a commutative algebra, so that differentiability and  analyticity can be well defined\cite{Davenport,Catoni2008} in the 4D realm, in close analogy with the (planar) complex analysis. 

The bicomplex algebra exhibits yet specific features in comparison with the complex algebra. Indeed, because of the presence of three versors, there exist \cite{Catoni2008} three principal conjugations of $q$ denoted by
\begin{eqnarray}
q^{\ast i}&=& z_{1}^{\ast}+jz_{2}^{\ast} , \\
q^{\ast j} &=& z_1-jz_2, \\
q^{\ast ij} &=& z_1^{\ast} - jz_{2}^{\ast}.
\end{eqnarray}
It is worth stressing that the bicomplex numbers generate then a non-Euclidian geometry (this could be already guessed from  the presence of the hyperbolic unit related to space-time geometry). In particular, the modulus, which is an invariant quantity of the geometry, necessarily \cite{Catoni2008} reads $\|q\|=\sqrt[4]{q q^{\ast i} q^{\ast j} q^{\ast ij}}=\sqrt{|z_1-z_2||z_1+z_2|}$. More interestingly, the geometry of the bicomplex space can be better grasped in a polar representation, where its peculiar topology becomes more explicit:
\begin{eqnarray}
q=\|q\| \, \exp\left(i \theta_{i}+j \theta_{j} +ij \theta_{ij} \right) \label{polar}
\end{eqnarray}
with the three real angles (or phases)
\begin{eqnarray}
\theta_{i} &=& \frac{1}{2} \left[ \arctan \frac{y_1+y_2}{x_1+x_2}+\arctan \frac{y_1-y_2}{x_1-x_2} \right],\\
\theta_{ij} &=&  \frac{1}{2} \left[ \arctan \frac{y_1+y_2}{x_1+x_2}-\arctan \frac{y_1-y_2}{x_1-x_2} \right], \\
\theta_{j} &=& \frac{1}{2} \ln \left|\frac{z_1+z_2}{z_1-z_2}\right|
.
\end{eqnarray}
This conformal mapping from the cartesian coordinates to the polar coordinates clearly highlights singular regions of the 4D space where the modulus vanishes and at least one of the three angles become undefined. These (forbidden) regions corresponding to the planes of equations $z_1 \pm z_2 =0$ concentrate all zero divisors\cite{zerodivisor} of the bicomplex algebra and can be regarded as forming a null cone.  Therefore, the modulus $\|q\|$ is a measure of nothing else but the distance to the null cone. Note that  two angles $\theta_i$ and $\theta_{ij}$ are circular as a result of their coupling to the imaginary unit $i$, whereas the third angle $\theta_j$ is, in contrast, hyperbolic.

\begin{figure}[t]
\centering
\includegraphics[width=0.35\textwidth]{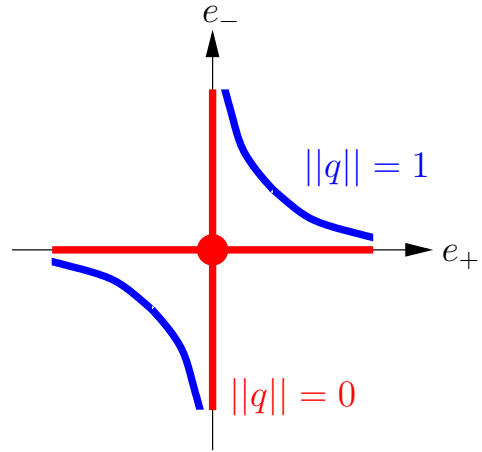}
\caption{(Color online) Schematic illustration of a 4D vortex according to the polar decomposition \eqref{polar}. The orthogonal axes which represent the axis planes (defined by the idempotent elements $e_{\pm}$) determine the location of the points of the 4D space where $\|q\|=0$. As for the product of two independent 2D vortices, these regions (of dimension 2) concentrate all the phase singularities. The distinguishing feature of the 4D vortex lies in the existence of a most singular point (corresponding to the intersection of the axis planes) where the hyperbolic phase $\theta_j$ characterizing the correlations between the two complex planes becomes additionally indeterminate. The hyperbola set by the consideration of a fixed nonzero $\|q\|$ illustrates the non-Euclidian nature of the bicomplex geometry.}
\label{fig}
\end{figure}

From a practical perspective, it appears very convenient to introduce  the idempotent elements $e_{\pm}=(1 \pm j)/2$, obeying the relations  $e_{\pm}^2=e_{\pm}$, $e_{+}e_{-}=0$ and $e_{+}+e_{-}=1$. Any bicomplex number can then be alternatively decomposed \cite{Davenport,Catoni2008} along the elements $e_{\pm}$, which can be seen as orthogonal axis planes, as $q=(z_1+z_2)e_{+}+(z_1-z_2)e_{-}$.
Within this idempotent projection into a pair of complex numbers, the invertible  bicomplex numbers (which can be alternatively decomposed in the polar representation) are all characterized by two nonzero components, i.e. by nonzero modulus $\|q \|$, since the singular regions precisely coincide with these axis planes defined by $e_{\pm}$. Note that the origin, which is the only point shared by the two  planes, turns out to be the most singular point since it corresponds to the unique place in the 4D space where the three angles appearing in the polar representation \eqref{polar} get simultaneously undefined (see Fig. \ref{fig} for a schematic illustration of the corresponding topological defect).

Being better familiar with the bicomplex algebra, we now come back to our physical problem. We first rewrite the
kinetic Hamiltonian \eqref{H2e} in terms of the bicomplex variable $q$ and its conjugations:
\begin{eqnarray}
\hat{H}_{0}^{2e}({\bf r})= \hat{H}(q,q^{\ast i}) + \left[ \hat{H}(q,q^{\ast i}) \right]^{\ast j}, \label{H2ebis}
\end{eqnarray}
where 
\begin{eqnarray}
\hat{H}(q,q^{\ast i})
=
\frac{\hbar \omega_c}{2} \left[-8 l_B^2 \partial_{q} \partial_{q^{\ast i}}+q \partial_q-q^{\ast i} \partial_{q^{\ast i}}+\frac{qq^{\ast i}}{8 l_B^2} \right]. \nonumber \\
\end{eqnarray}
Then, it can be straightforwardly checked that the following bicomplex-valued wave functions 
\begin{eqnarray}
  \langle {\bf r} |n,m,\lambda,{\bf R} \rangle =\frac{1}{\sqrt{2}} \left(\begin{array}{c} \Psi_{n,m,{\bf R}}({\bf r}) \\  \lambda \Psi^{\ast j}_{n,m,{\bf R}}({\bf r}) \end{array} \right) \label{vor4D}
\end{eqnarray}
with
\begin{eqnarray}
\Psi_{n,m,{\bf R}}({\bf r})=\frac{(2\pi l_B^2)^{-1}}{\sqrt{n! \,m!}} \left( \frac{q-Q}{2 l_B}\right)^{n} \left( \frac{q^{\ast ij}-Q^{\ast ij}}{2 l_B}\right)^{m} \nonumber \\
\times \, e^{-\frac{|z_1|^2+|Z_1|^2-2Z_1 z_1^{\ast}}{4 l_B^2}} \, e^{-\frac{|z_2|^2+|Z_2|^2-2Z_2 z_2^{\ast}}{4 l_B^2}}, \hspace*{0.5cm}
\end{eqnarray}
are eigenstates of the two-electron kinetic energy operator $\hat{H}_0^{2e}(\mathbf{r})$ in the symmetrical gauge, with the energy quantization $E_n=(n+1)\hbar \omega_c$. 
Here $n$ and $m$ are positive integers and $\lambda=\pm 1$ is a band index. Analogously to the definition of $q$, the bicomplex number $Q$ reads in terms of the complex components $Z_1=X_1+iY_1$ and $Z_2=X_2+iY_2$ as $Q=Z_1+jZ_2$. The quantum numbers ${\bf R}=({\bf R}_1,{\bf R}_2)$ characterize the arbitrary position  of the origin of the bicomplex frame in the 4D space. Therefore, as in the case of the 2D complex vortex states, these quantum numbers are somehow defined from a translation operation (yet here in a non-Euclidian geometry), which is a common way to generate coherent states character.
The spinorial structure of the states \eqref{vor4D} is required  in order to give a satisfactory interpretation of the square modulus as a probability density, see the discussion in the next section. 
It also naturally results from the symmetry of the kinetic energy operator \eqref{H2ebis} with respect to the hyperbolic conjugation (operation $^{\ast j}$), which produces an additional two-fold degeneracy (represented by the quantum number $\lambda$) for the pair Landau levels when $(n,m) \neq (0,0)$. 

One should be aware that the polar decomposition \eqref{polar}, which is the best representation of the bicomplex geometry, is always implicitly assumed: when $(n,m) \neq (0,0)$, the functions \eqref{vor4D} not only vanish  for  ${\bf r}={\bf R}$ as one may naively guess at first sight, but also for electronic positions ${\bf r}$ lying in the whole null cone, \textit{i.e.}, here when $(z_1-Z_1) + (z_2-Z_2)=0$ or when $(z_1-Z_1) - (z_2-Z_2)=0$. The fact that these regions of singularities for the phases of the wave function are of codimension 2 is naturally expected for vortex-like defects in the 4D space (this was for example already the case when considering the product states of two independent 2D vortices). It is important to notice that the polynomial part of the states  \eqref{vor4D} actually represents a kind of gravitation-like distortion which structures the whole 4D space.   It describes bicomplex harmonic functions consisting of four real components of four variables, which are very tightly linked [each  of the four real components defined in the basis $(1,i,j,ij)$ obeys a 4D Laplace's equation].

The two non-negative integers $n$ and $m$, which stem from the presence of two circular angles in the bicomplex geometry, are nothing but the  winding numbers for the pair motion of electrons enclosing the singular axis planes. 
An illustration of the genuinely 4D nature of this quantum pair cyclotron motion, which can not be reduced to two independent 2D electronic orbital motions for topological grounds, is that only the winding number $n$ now contributes to the quantization of the two-electron kinetic energy. The other winding number $m$, which always appears in association with the usual conjugation $^{\ast i}$ of complex numbers, plays the role of a negative circulation, hinting at a semi-classical orbital motion of the correlated electrons with an axis of rotation pointing in the opposite direction to the applied magnetic field. 

Although the 2D vortex states \eqref{vor} and their 4D counterparts \eqref{vor4D} look quite similar, the major difference between them can be inherently found in the different space topologies that they embed. In both cases, the nontrivial topologies of the complex and bicomplex algebras  give rise to quantum eigenstates containing long range structures and characterized by robust quantum numbers (\textit{i.e.} topological attributes).
The bicomplex representation of the eigenstates of the two-electron kinetic energy operator  has the additional peculiarity of also relaying some long-range structure to the electronic pair correlations, given that the electronic coordinates ${\bf r}_1$ and ${\bf r}_2$ are intrinsically intertwined. As in the one-electron case, the robustness of the 4D vortex states can only be established within the full process of Landau level degeneracy lifting. At the two-electron level, the electron-electron interaction  sets an intermediate energy scale into the high magnetic field problem, which is expected to give rise to some gap protection for the discrete pair degeneracy quantum numbers. Additionally,  the  smooth contributions to the potential energy vindicate the use of a semi-classical treatment for the description of the slow components of the electron pair motion, similarly to the single electron case. 
 
Finally, it should be noted that within the bicomplex algebra one literally realizes a complete fusion of the individual electronic cyclotron motions via a novel kind of topological entanglement, which is formally allowed from the viewpoint of the Schr\"{o}dinger's equation alone. 
The main issue, which will be  only partially settled in the remainder of this paper, is whether this bicomplex representation of the electronic states  conforms to some physical reality at high magnetic fields.

\section{Completeness relation and Hilbert space of pair vortex states \label{basis}}

In this section, we aim at arguing and  proving that the set of 4D vortex eigenstates $|n,m,\lambda, {\bf R} \rangle$ forms a reliable representation of the electronic quantum states. An obvious concern is that going beyond complex numbers may cause problems in the physical interpretation\cite{Adler1995}.  In order to preserve the standard structure  of  quantum mechanics, it turns out important to keep the inner product between two arbitrary states $|\Psi_1 \rangle$ and $|\Psi_2 \rangle$ of the space of squared integrable functions, which is defined only in terms of the conjugation $^{\ast i}$ (with respect to the imaginary unit $i$ of complex numbers) as
\begin{eqnarray}
\langle \Psi_1 | \Psi_2 \rangle=\int d^4 {\bf r} \, \Psi_1^{\ast i}({\bf r}) \Psi_2({\bf r}), \label{prod}
\end{eqnarray}
where we have used the standard bra-ket notation. When considering eigenfunctions \eqref{vor4D}, the product \eqref{prod} is now defined in a bicomplex functional space with the target space being the bicomplex numbers space. Most importantly, it still constitutes a scalar quantity which is invariant under all possible rotational transformations of the coordinate system. Indeed, when inserting expressions \eqref{vor4D} into  \eqref{prod}, we generically deal with the spatial integration of the bicomplex quantity $q_1 q_{2}^{\ast i} q_{3}^{\ast j} q_{4}^{\ast ij}$ (with $q_1$, $q_2$, $q_3$ and $q_4$ representing four different bicomplex numbers), which constitutes an invariant form of degree 4 for the bicomplex metric.

It can be easily checked that the states  $|n,m,\lambda, {\bf R} \rangle$ are normalized according to the standard definition \eqref{prod} for the inner product. As for 2D complex vortex states, two different pair vortex eigenstates are expected to have a non-orthogonal overlap by virtue of their coherent-state character. After calculations (see Appendix \ref{apOver}), we get the result

% overlap between vortex states

\begin{eqnarray}
\langle n,m,\lambda, {\bf R} | n',m',\lambda',{\bf R}' \rangle = \delta_{n,n'} \, \langle {\bf R}_1 |{\bf R}'_1 \rangle \, \langle {\bf R}_2 |{\bf R}'_2 \rangle \nonumber \\
\times \frac{1}{2} \left[ \gamma_{m;m'}({\bf R}'-{\bf R})+ \lambda \lambda' \gamma_{m;m'}^{\ast^j}({\bf R}'-{\bf R}) \right],\hspace*{0.5cm} \label{over}
\end{eqnarray}
with
\begin{eqnarray}
\gamma_{m;m'}({\bf R}) =\sum_{p,p'} \delta_{p,p'} \left( \begin{array}{c} m \\ p \end{array} \right)  \left( \begin{array}{c} m' \\ p' \end{array} \right) \sqrt{\frac{p! \, p'!}{m! \, m'!}} \nonumber  \\ \times  \left(\frac{Q^{\ast^j}}{2l_B} \right)^{m-p} \left(-\frac{Q^{\ast^{ij}}}{2l_B} \right)^{m'-p'}. \label{gamma}
\end{eqnarray}
In addition to the typical Gaussian overlap of coherent states, the overlap function \eqref{over} displays a polynomial bicomplex dependence on ${\bf R}'-{\bf R}$ via the quantity $\gamma_{m;m'}({\bf R}'-{\bf R})$. As a result, it is not only non-orthogonal with respect to the quantum numbers ${\bf R}$, but also with respect to the other degeneracy quantum numbers $m$ and $\lambda$. 
 Interestingly, if we now consider coinciding vortex positions ${\bf R}={\bf R}'$, the overlap \eqref{over} then becomes entirely  diagonal with respect to the discrete quantum numbers

\begin{eqnarray}
\langle n,m,\lambda, {\bf R} | n',m',\lambda',{\bf R} \rangle = \delta_{n,n'} \, \delta_{m,m'} \, \delta_{\lambda,\lambda'}.
\end{eqnarray}
This relation brings back some symmetry between the positive circulation quantum number $n$ and the negative circulation quantum number $m$, which was apparent at the level of the wave functions \eqref{vor4D}.
However, because of its original intricate coupling with the vortex position ${\bf R}$ [see Eq. \eqref{over}], it is expected that  $m$ plays a strikingly different role from the Landau level index $n$ in the effective dynamics of the 4D vortex.

% completeness relation

As it could be anticipated from their definition in terms of harmonic modes of the 4D bicomplex geometry, the set of states $|n,m,{\bf R},\lambda \rangle $ obeys a closure relation (see proof in Appendix \ref{apComp}), with a form relatively similar to that of relation \eqref{comp}:

\begin{eqnarray}
 \int \!\! \frac{d^4 {\bf R}}{(2 \pi l_B^2)^2} \!\! \sum_{n,m,\lambda} f(m) \, |n,m,\lambda,{\bf R}\rangle \langle n,m, \lambda,{\bf R} |= 1\!\!1_2,
%\left( \begin{array}{cc}1 & 0 \\ 0 & 1 \end{array}\right)
\hspace*{0.5cm} \label{comp4D}
\end{eqnarray}
where $1\!\!1_2 = 1\!\!1 \otimes 1\!\!1_\lambda$.
This identity reveals some flexibility  of the representation (in addition to that already provided by the doubly continuous character of ${\bf R}$), since the weight function $f(m)$ in Eq. \eqref{comp4D} is not yet entirely determined and obeys the sole constraint
\begin{eqnarray}
\sum_{m=0}^{+\infty} f(m)=1. \label{constraint}
\end{eqnarray}
The simplest choice is to set the quantum number $m$ to a fixed value so that it is not required to sum over all positive integers $m$ in Eq. \eqref{comp4D}. The integer $m$ somehow acquires the status of a good ``extra'' quantum number since then only the diagonal matrix elements \eqref{over} with $m=m'$ become necessary in the quantum representation.
In this case, the algebra with respect to the vortex positions ${\bf R}$ becomes pretty analogous to that of the so-called generalized coherent states\cite{Philbin}.

% Hilbert space

The resolution of unity \eqref{comp4D} considered for a fixed given $m$ shows that the collection of vortex states  \eqref{vor4D} spans a Hilbert space, whose any state can be
expressed as the linear  combination
\begin{eqnarray}
\left| \Psi \right. \rangle = \!\! \int \!\! \frac{d^4 {\bf R}}{(2 \pi l_B^2)^2} \! \sum_{n,\lambda} e^{-\frac{{\bf R}_1^2+{\bf R}_2^2}{4l_B^2}}  \left( \begin{array}{c}  c_{n,m}({\bf R}) \, |n,m, {\bf R}\rangle \\  \lambda c^{\ast j}_{n,m}({\bf R}) \, |n,m, {\bf R}\rangle^{\ast j} \end{array} \right), \nonumber \\ \label{combi}
\end{eqnarray}
with coefficients taking the following form $c_{n,m}({\bf R})=(2l_B^2 \partial_{Q^{\ast ij}}-Q^{\ast j})^m \, c_{n,m}(Q^{\ast ij})  $ where $c_{n,m}(Q^{\ast ij})$ are analytical functions of the variables $Q^{\ast ij}$.
%Here the elements  $c_{n,m,\lambda}({\bf R})$ of the base field are taken to be complex numbers  as in standard quantum mechanics. This means that the vector space remains complex, while the basis vectors have been originally defined   in terms of the electronic variables by use of the bicomplex algebra.  
Note  that the norm of any state belonging to this Hilbert space with respect to the inner product \eqref{prod} is real and positive, since from Eqs. \eqref{over} and \eqref{combi} we get after straightforward calculations
\begin{eqnarray}
\langle \Psi | \Psi \rangle &=& \int \!\! \frac{d^4 {\bf R}}{(2 \pi l_B^2)^2} \! \sum_{n}  2\, \exp \left[-\frac{{\bf R}_1^2+{\bf R}_2^2}{2l_B^2} \right]   \nonumber \\  &&
\times \left[c_{n,m}({\bf R})c^{\ast i}_{n,m}({\bf R}) +c_{n,m}^{\ast j}({\bf R})c^{\ast ij}_{n,m}({\bf R}) \right]. \hspace*{0.6cm} \label{PsiPsi}
\end{eqnarray}
Therefore,  the modulus squared of the probability amplitude can still be physically viewed as representing the electronic probability density.

% positive reality of eigenvalues

Finally, let us consider the general form of the matrix elements of the potential energy $V({\bf r}) $ in the pair vortex representation. Here $V(\mathbf{r})\equiv V({\bf r}_1,{\bf r}_2)$ incorporates the interactions between electrons  and the potentials due to impurities and electrostatic confinement seen by the two electrons forming the pair. It thus generically depends on both two-dimensional electronic coordinates ${\bf r}_1$ and ${\bf r}_2$. As a result of the spinorial form of the wave functions \eqref{vor4D}, it is straightforward to see that these matrix elements can be generally written as

\begin{eqnarray}
\langle n,m,\lambda, {\bf R} |\hat{V}| n',m',\lambda',{\bf R}' \rangle =  \frac{1}{2} \left[ \frac{}{} \langle n,m,{\bf R}|\hat{V}| n',m',{\bf R}' \rangle \nonumber\right.  \\ \left. + \lambda \lambda ' \left(\langle n,m,{\bf R}|\hat{V} |n',m',{\bf R}' \rangle \right)^{\ast j}
\right] .\hspace*{1cm} \label{mat}
\end{eqnarray}
Obviously, the purely hyperbolic terms, proportional to the unit $j$, correspond to off-diagonal contributions in the $2 \times 2$ $\lambda$-basis, while diagonal blocks (for $\lambda=\lambda'$) are complex numbers. This construction guarantees the real character of the energy spectrum, since the Hamiltonian operator is then represented by an Hermitian matrix with respect to both the usual conjugation and the hyperbolic conjugation.
From simple calculations (see Appendix \ref{apOver}), it can be shown that the quantity $\langle n,m,{\bf R} |\hat{V}| n',m',{\bf R}' \rangle$ in Eq. \eqref{mat} takes the generic form
\begin{eqnarray}
\langle n,m,{\bf R} |\hat{V}| n',m',{\bf R}' \rangle =  \langle {\bf R}_1 |{\bf R}'_1 \rangle \, \langle {\bf R}_2 |{\bf R}'_2 \rangle \sum_{p,p'} \sqrt{\frac{p! \, p'!}{m! \, m'!}} \,\,\,\,\,\,  \nonumber \\
\times  \! \left( \begin{array}{c} m \\ p  \end{array}\right) \! \left( \begin{array}{c} m' \\ p'  \end{array}\right) \!\left(\delta Q^{\ast j}\right)^{m-p} \! \left(-\delta Q^{\ast ij}\right)^{m'-p'} \!\!\!   v_{n,p;n',p'}(\tilde{{\bf R}}), \nonumber \\
 \label{vmat}
\end{eqnarray}
where $\delta Q =(Q'-Q)/(2 l_B)$ and $\tilde{{\bf R}}=(\tilde{{\bf R}}_1,\tilde{{\bf R}}_2)$ is defined through an analytical continuation of the vortex position ${\bf R}=({\bf R}_1,{\bf R}_2) $ to the complex plane, such that $\tilde{{\bf R}}_{s}=\left[ {\bf R}_s+{\bf R}'_s+i({\bf R}'_s-{\bf R}_s) \times \hat{{\bf z}}\right]/2$ for $s=1,2$. Importantly, we recover within expression \eqref{vmat} a generic property\cite{Sudarshan} of the coherent-state representation, namely, that any operator or state are uniquely determined by their diagonal matrix elements. Indeed, the entire information contained in the potential matrix elements can be actually found when considering the latter at coinciding vortex positions ${\bf R}={\bf R}'$, within its reduced matrix elements $v_{n,p;n',p'}({\bf R})$, which explicitly read
\begin{eqnarray}
v_{n,p;n',p'}({\bf R})&=&\frac{1}{\sqrt{n!\,n'! \, p! \, p'!}} \int \!\!\! \frac{d^{4} {\bf r}}{\pi^2/4} e^{-2 {\bf r}^2}  V({\bf R}+2 l_B {\bf r}) \nonumber \\ &&
\times (q)^{n'} (q^{\ast i})^{n} (q^{\ast j})^{p} (q^{\ast ij})^{p'}.\label{reducedv}
\end{eqnarray}
For a constant potential $\hat{V}=1\!\!1$, we immediately  find again the above expression \eqref{over} for the overlap between two arbitrary pair vortex states, since then $v_{n,p;n',p'}({\bf R})=\delta_{n,n'} \, \delta_{p,p'}$. Eqs. \eqref{mat} and \eqref{reducedv} constitute the basic elements of an effective potential seen by the pair vortex which will appear in the equation of the motion. Because it involves non-trivial technical developments beyond the scope of the present paper, the mathematical derivation of an equation of motion for the pair vortex in the presence of both a disorder potential and electron-electron interactions is postponed to future work. In the next section, we shall, however, anticipate the effects of the latter in a purely qualitative way.

\section{Perspectives for the fractional quantum Hall effect \label{pers}}

So far, we have only discussed the peculiar form and properties of a subset of quantum states  for two non-interacting (yet correlated) electrons in an external perpendicular magnetic field. We now analyze at a qualitative level the degeneracy lifting process of the pair Landau level by both one-body and two-body interaction potentials at the light of the vortex representation of the  states previously developed. Obviously, the quest of good quantum numbers at the quantum mechanical level underlies in a way the choice of the most relevant decomposition of the global motion of the electron pair into elementary motions. This relevance manifests through the presence of a hierarchy of energy scales (or time scales) for  these elementary motions, which will be exploited in order to carry out a separation between the fast and slow degrees of freedom. 

%decomposition of motion = separation of energy scales

Clearly, the cyclotron motions of the electrons associated with the Landau level quantum number set the highest energy scale of this hierarchy at high magnetic fields. For two independent  electrons in circular motion one naturally expects the presence of two integral quantum numbers. However, we have seen that the Schr\"{o}dinger's equation also allows, in principle, bicomplex wave solutions describing correlated cyclotron motions and contributing to the kinetic energy quantization with a single pair Landau level index. The byproduct of this fast correlated rotation is that the  second integral quantum number characterizing a counter-propagating orbital pair motion is then relegated to the kinetic energy degeneracy. 

This integral degeneracy, as well as the degeneracy with respect to the relative guiding center ${\bf R}_1-{\bf R}_2$, gets intrinsically lifted by the interactions between the two electrons forming the pair which provide a sub-leading energy scale, since at high magnetic fields the effective Coulomb interaction (integrated over the cyclotron orbit) scales with the square root of the magnetic field. Classically, the central interaction potential  imparts at high magnetic fields a relatively fast rotational motion of the guiding centers ${\bf R}_1$ and ${\bf R}_2$ around each other. In the quantum case, this periodic motion gives rise to  bound states\cite{Girvin1983}, which are best characterized by another discrete good quantum number (a quantized relative angular momentum) in place of the quantity ${\bf R}_1-{\bf R}_2$. Thus, our 4D vortex approach suggests the existence of gaps for the pair energies which are labeled by the collection of two different integral quantum numbers (a dependence of the gaps on the band quantum number $\lambda=\pm 1$ is also expected) in addition to the Landau level index. Finally, the lowest energy scale is associated with the  motion of the  center of mass of the two guiding centers ${\bf R}_1$ and ${\bf R}_2$, which is  induced by the smooth part of the pair potential energy due to impurities and electrostatic confinement. This slow motion can be well captured at finite temperatures by semiclassical approximations which would lead to a smooth dispersion of the effective pair energy with respect to the center of mass $({\bf R}_1+{\bf R}_2)/2$.

% why pairs are relevant degree of freedom?

In many situations the consideration of the two-electron problem turns out to be very instructive \cite{Girvin2004} in order to grasp crucial parts of the physics of the many-electron system or at least to develop a feel for it. For instance, the celebrated Laughlin's wave functions \cite{Laughlin1983}, which successfully describe the sequence of some peculiar fractions of  the fractional quantum Hall effect, can be understood \cite{Girvin2004} as a generalization of the two-particle states, especially highlighting pairwise correlations between the electrons. The fact that a pair of electrons in the presence of a repulsive interaction potential has a discrete spectrum is a central feature underlying the existence of excitation gaps in the many-electron problem. This result  mostly stems from the severe restrictions  imposed by the magnetic field on the quantum states after projection on the lowest Landau level.  This projection usually amounts to get rid of the electronic cyclotron motion degrees of freedom, so that the correlations essentially take place through the interactions between the guiding center degrees of freedom. One key point considered in this paper, which is discarded in the usual treatment of the two-electron problem, is to introduce into the description pair correlations between the two electronic cyclotron motions already at the level of the Landau level quantization process, \textit{i.e.}, before freezing the kinetic energy. 
As a result,  the lowest Landau level projection for the two-electron problem amounts to get rid of only one degree of freedom instead of two when restricting solutions of the Schr\"{o}dinger's equation to analytical complex-valued wave functions.

%link with composite fermions?
It is then very tempting to relate the different fractions observed in the fractional quantum Hall effect, which are characterized by at least two independent integers,
to the presence of two good discrete quantum numbers for the pair motion, namely, the relative guiding center angular momentum and 
the discrete negative circulation released through the generation of cyclotron quantum correlations. Indeed, Jain's sequence of fractions \cite{Jain1989,Jain2000,Jain2007} showing the most pronounced features in the transport properties also involves two independent integers. The interpretation of the quantum states is usually grasped by means of the concept of weakly interacting composite fermions (presented as electrons bound to an even number of flux quanta) moving in a reduced magnetic field \cite{Jain2000}. Theoretically, the corresponding ground states are derived \cite{Jain2007} as a generalization of the Laughlin's trial wave functions by involving  the contributions of higher Landau levels before projecting again on the lowest Landau level, a construction which  appears rather bizarre from the physical point of view at high magnetic fields. It is worth noting that, within the pair vortex representation of the quantum states, the negative circulation is expected to manifest as an effective kinetic energy, since it is originally associated with the electronic orbital motion. In some sense, it thus causes similar effects to those originating from the composite fermion phenomenological construction. One may easily envision that the quantum counter-rotation of the electrons embodied by the negative circulation quantum number amounts to a screening of the original magnetic field. In contrast to the one-body case, the pair vortex thus sees an effective magnetic field in addition to an effective potential energy, as a result of the integration  over the fast (frozen) orbital degrees of freedom.

The fruitful understanding of the fractional quantum Hall effect as an integral quantum Hall effect of weakly interacting composite particles which has been obtained during the past decades \cite{Jain2007} definitely calls for a unifying microscopic principle. For the sake of consistency, this common physical guideline should be found again at the core of the theoretical microscopic treatments of both the integral and fractional quantum Hall effects. At the beginning of this paper we have insisted on the existence of vortex-like solutions to the Schr\"{o}dinger's equation at the level of the one-body kinetic energy operator  \eqref{Ham1el} in a magnetic field. Therefore, the (integer) Landau gap responsible for the integer quantum Hall effect can be fundamentally understood as the signature of a topological defect. Then, we have shown that the two-electron kinetic energy operator also embeds peculiar solutions representing the free pair motion as a 4D vortex, again linking the Landau level index with a quantized positive circulation. Similarly to the one-electron case, the creation of this topological object provides a microscopic  mechanism that sustains the presence of energy gaps in the collective electronic modes at high magnetic fields.

A direct objective for future work is the derivation of the interaction energy gaps for the electron pair vortices, which should reveal specific dependencies on the magnetic field and on the discrete pair quantum numbers that could be compared with experiments. An important step towards the construction of a microscopic theory for the fractional quantum Hall effect will be to specify the connection between the many-electron problem and an effective problem formulated in terms of 4D vortices of electron pairs.
It will also remain to clarify the role of  quantum statistics and the fundamental mechanism responsible for the fractionalization of the Hall conductance in the effective problem. 
%If the phenomenological mapping between the integer and the fractional quantum Hall effects is to be taken literally at the microscopic level,  Fermi statistics is expected to play a crucial role essentially in setting the occupation of the energy levels, but not within the process of quantization of the energy spectrum of the electron pairs.
Finally,  Green's  function techniques such as those being already developed\cite{Champel2008,Champel2009,Champel2010} for the integer quantum Hall regime  should also provide microscopic derivations of the transport coefficients in the strongly correlated regime within a semiclassical framework exploiting the slow character of the center of mass pair vortex motion.

\section{Conclusion \label{conc}}

In summary, we have highlighted the existence of singular microscopic quantum solutions for the cyclotronic motion of electrons, which provide a topological and hydrodynamic view in order to capture the effects on these states induced by the presence of arbitrary smooth disorder and interactions between electrons. To that purpose, we have principally generalized in this paper this vortex approach to the two-body problem by pinpointing a peculiar subset of pair vortex coherent states which embody a topologically-protected entanglement of the two electronic orbital motions. The corresponding correlations between the two electrons are then intrinsically long-range and built in through the non-Euclidean geometry of the bicomplex numbers, which generalize to the 4D space the concept of complex numbers. We have also put forward that the introduction of the bicomplex algebra does not fundamentally alter the standard formulation and interpretation of quantum mechanics, in so far as the usual scalar product functional form is kept invariant. % and the vector space remains complex. 
Importantly, we have shown that the set of pair vortex eigenstates can be used as a reliable representation of the electronic states, since it forms an overcomplete basis of an enlarged Hilbert space. Finally, we have addressed qualitatively the problem of the degeneracy lifting process of the pair Landau levels by both strong electron-electron interactions and smooth disorder, and have argued that the pair vortex motion in the lowest Landau level is expected to be  characterized by energy gaps labeled by two good discrete (independent) quantum numbers. %Because the 4D vortices of electron pairs are expected to feel an effective magnetic field, we have speculated a possible relationship between the presence of these two integers and Jain's sequence of fractions for the Hall conductance in the fractional quantum Hall regime.

%, we have  developed at the qualitative level a preliminary scenario in terms of weakly interacting electron pairs which could allow for a global microscopic understanding of the fractional quantum Hall regime accounting for both the effects of strong electron-electron interactions and disorder.

%One may envision generalization of the algebra to account for three-particle quantum correlations. 

%More complicated singularities can be envisioned in quantum mechanics by going beyond complex wave functions. 

\section*{Acknowledgments}

Interesting discussions with D. Basko, V. Rossetto and D. Spehner are gratefully acknowledged. D.H.-P. was supported by the RTRA Nanosciences Foundation in Grenoble.

\vspace*{0.5cm} 
\appendix

\section{Overlap and potential matrix elements \label{apOver}}

In this Appendix, we detail the derivation of the general expression \eqref{vmat} from Eq. \eqref{mat}. By definition, the matrix elements of the potential energy $V(\mathbf{r})$ read, in terms of the pair vortex wave functions,
\begin{eqnarray}
\langle n,m, {\bf R} |\hat{V}|n',m',{\bf R}' \rangle =\int d^{4}{\bf r} \, \Psi^{\ast i}_{n,m,{\bf R}}({\bf r})  V(\mathbf{r}) \Psi_{n',m',{\bf R}'}({\bf r}). \nonumber \\
\end{eqnarray}
We first introduce the new integration variables $\tilde{{\bf r}}=({\bf r}-\tilde{{\bf R}})/2l_B$ with $\tilde{{\bf R}}=(\tilde{{\bf R}}_1,\tilde{{\bf R}}_2)$ and $\tilde{{\bf R}}_{s}=[{\bf R}_s+{\bf R}'_s+i({\bf R}'_s-{\bf R}_s) \times \hat{{\bf z}}]/2$ for $s=1,2$, what amounts to shift the four Cartesian coordinates $(x_1,y_1,x_2,y_2)$ composing the 4D vector ${\bf r}$ by constant complex quantities.
Note that the four corresponding contours are deformed to the real axes thanks to the analyticity of the integrated functions.  After this shift, the overlap then reads
\begin{eqnarray}
\langle n,m, {\bf R} |\hat{V}|n',m',{\bf R}' \rangle =\langle {\bf R}_1| {\bf R}'_1 \rangle \, \langle {\bf R}_2| {\bf R}'_2 \rangle 
\int \!\! \frac{d^{4} \tilde{{\bf r}} }{\pi^2/4} e^{-2 \tilde{{\bf r}}^2} \,\,\,  \nonumber \\
\times
\left( 
\tilde{q}\right)^{n'} \!\! \left( \tilde{q}^{\ast i}\right)^n  \!\! \left( \tilde{q}^{\ast j}+\delta Q^{\ast j}\right)^m  \!\! \left( \tilde{q}^{\ast ij}-\delta Q^{\ast ij}\right)^{m'} \!\! V(\tilde{{\bf R}}+2 l_B \tilde{{\bf r}}), \nonumber \\
\end{eqnarray}
where $\delta Q=(Q'-Q)/2 l_B$.
Expanding the two polynomial parts shifted by the quantity $\delta Q$ by using the binomial formula,  we straightforwardly get the result written in Eqs. \eqref{vmat} and \eqref{reducedv}. In general, it turns out convenient not to specify explicitly the boundaries of the discrete sums, which can be accounted for in the binomial coefficients by using their extended definition 
\begin{eqnarray}
\left( \begin{array}{c} m \\ p \end{array}\right) = 0,
\end{eqnarray}
when either $p> m$ or $p <0$.
\vspace*{0.5cm}

\section{Proof of completeness relation \label{apComp}}

In this Appendix, we provide a proof of the closure relation \eqref{comp4D}. We first consider the quantity
\begin{eqnarray}
{\cal I}= \sum_{n,\lambda} \int \frac{d^4 {\bf R}}{(2 \pi l_B^2)^2} \langle n',m',\lambda',{\bf R}'| n,m,\lambda,{\bf R} \rangle \nonumber \\ 
\times \langle n,m,\lambda,{\bf R} |n'',m'',\lambda'',{\bf R}'' \rangle \label{calI}
\end{eqnarray}
defined for arbitrary  bra  $\langle n',m',\lambda',{\bf R}'|$ and ket $|n'',m'',\lambda'',{\bf R}'' \rangle$ states. Using expressions \eqref{over} and \eqref{gamma}, then shifting in the complex plane the variable of integration ${\bf R}$ in order to 
center the Gaussian exponential factors as done in Appendix \ref{apOver}, we get after summation over the discrete sums,
\begin{eqnarray}
{\cal I}=\frac{1}{2} \left[ {\cal G}^{\ast j}({\bf R}''-{\bf R}') + \lambda' \lambda'' {\cal G}({\bf R}''-{\bf R}')\right] \nonumber \\
\times  \langle {\bf R}'_1 | {\bf R}''_1 \rangle \langle {\bf R}'_2 | {\bf R}''_2 \rangle , \label{I}
\end{eqnarray}
where 
\begin{eqnarray}
{\cal G}({\bf R}')=\sum_{p_1,p_2} \left( \begin{array}{c} m' \\ p_1 \end{array} \right) \left( \begin{array}{c} m \\ p_1 \end{array} \right)  \left( \begin{array}{c} m \\ p_2 \end{array} \right) \left( \begin{array}{c} m'' \\ p_2 \end{array} \right) \frac{p_1!p_2!}{m! \sqrt{m'!m''!}} \nonumber \\
\times \int \!\! \frac{d^4 {\bf r}}{\pi^2/4} e^{-2{\bf r}^2} \left( q+\frac{Q'}{2 l_B}\right)^{m'-p_1}\left( -q^{\ast i}\right)^{m-p_1}\left( -q\right)^{m-p_2} \nonumber \\ 
\times \left( q^{\ast i}-\frac{Q^{\prime \ast i}}{2 l_B}\right)^{m''-p_2}. \hspace*{1cm}
\end{eqnarray}
Using twice the binomial formula such that 
\begin{eqnarray}
\left( q+\frac{Q'}{2 l_B}\right)^{m'-p_1} \!\! &=&\sum_{p}\! \left( \begin{array}{c} \!\! m'-p_1 \!\! \\ \!\! m'-p  \!\! \end{array} \right) \! \left(q \right)^{p-p_1} \!\!  \left( \frac{Q'}{2l_B}\right)^{m'-p}, \nonumber \\
\left( q^{\ast i}-\frac{Q^{\prime \ast i}}{2 l_B}\right)^{m''-p_2} \!\!\!\!\! &=&\sum_{p'} \! \left( \begin{array}{c} \!\! m''-p_2 \!\! \\  \!\! m''-p' \!\!\end{array} \right) \! \left(q^{\ast i} \right)^{p'-p_2} \!\! \left(\! -\frac{Q^{\prime \ast i}}{2l_B}\right)^{m''-p'}, \nonumber
\end{eqnarray}
and the identity
\begin{eqnarray}
\int \!\! \frac{d^4 {\bf r}}{\pi^2/4} e^{-2 {\bf r }^2} \! q^{n'} \! \left(q^{\ast i}\right)^{n} \!\left(q^{\ast j}\right)^{m} \! \left(q^{\ast ij}\right)^{m'} \! =n! m! \, \delta_{n,n'} \delta_{m,m'}, \nonumber
\end{eqnarray}
which can be easily established by different means within the bicomplex algebra, we  obtain
\begin{eqnarray}
{\cal G}({\bf R}')= \sum_{p,p'} \delta_{p,p'} \left( \begin{array}{c} m' \\ p \end{array} \right) \left( \begin{array}{c} m'' \\ p' \end{array} \right) \sqrt {\frac{p! \, p'!}{m'! \, m''!}} \nonumber \\
\times \left(\frac{Q'}{2l_B} \right)^{m'-p} \left(-\frac{Q^{\prime \ast i}}{2l_B} \right)^{m''-p'} \, g_{p,m}, \label{g}
\end{eqnarray}
where we have used 
\begin{eqnarray}
\left( \begin{array}{c} m' \\ p_1 \end{array} \right)\left( \begin{array}{c} m'-p_1 \\ m'-p \end{array} \right)& =&\left( \begin{array}{c} m' \\ p \end{array} \right)\left( \begin{array}{c} p \\ p_1 \end{array} \right), \\
\left( \begin{array}{c} m'' \\ p_2 \end{array} \right)\left( \begin{array}{c} m''-p_2 \\ m''-p' \end{array} \right)&=&\left( \begin{array}{c} m'' \\ p' \end{array} \right)\left( \begin{array}{c} p' \\ p_2 \end{array} \right).
\end{eqnarray}

We can note from Eqs. \eqref{I} and \eqref{g} that the quantity ${\cal I}$ takes at this point almost the form of the overlap \eqref{over} between two pair vortex states up to the coefficient $g_{p,m}$
which is given by
\begin{eqnarray}
g_{p,m}=\sum_{p_1,p_2} (-1)^{p_1+p_2} \frac{p_1! \, p_2!}{p! \, m!} \left( \begin{array}{c} m \\ p_1 \end{array} \right) \left( \begin{array}{c} m \\ p_2 \end{array} \right)\left( \begin{array}{c} p \\ p_1 \end{array} \right) \left( \begin{array}{c} p \\ p_2 \end{array} \right) \nonumber \\
\times (m+p-p_1-p_2)! . \hspace*{1cm} \label{coefg}
\end{eqnarray}
Actually, this coefficient $g$ turns out be equal to unity irrespective of the integers $p$ and $m$. This comes out rather straightforwardly from the following identity \cite{Ruiz}
\begin{eqnarray}
\sum_{p} (-1)^{p}  \left( \begin{array}{c} m \\ p \end{array} \right) P(x+m-p)=m! \, a_{m}, \label{sumRuiz}
\end{eqnarray}
which holds for any polynomial $P(x)$ of degree less than or equal to $m$ (here $a_m$ is the coefficient of degree $m$ of $P$). In particular, if the polynomial $P$ is of degree strictly less than $m$, this means that the result of summation in \eqref{sumRuiz} amounts to zero, since then $a_m=0$. The above identity can be easily understood by recognizing the forward difference operator $D_x$ applied $m$ times to the polynomial $P(x)$. This linear operator in the space of polynomials reduces the degree of any polynomial function by one and it is therefore understood as a discrete version of the derivative representing quasi-locality. Indeed, we have $D_x P(x)=P(x+1)-P(x)$, and 
\begin{eqnarray}
\left(D_x \right)^m P(x)=\sum_{p} (-1)^{p}  \left( \begin{array}{c} m \\ p \end{array} \right) P(x+m-p).
\end{eqnarray}
The different factorials appearing in Eq. \eqref{coefg} can be gathered together and seen  as a polynomial of some given variable $x$, in such a way that the sum over one of the two integers $p_1$ or $p_2$ can be first performed by use of identity \eqref{sumRuiz}. As a consequence, only one term of the  second sum remains nonzero from the polynomial degree constraint and yields  the result $g_{p,m}=1$.

Finally, in order to get the completeness relation \eqref{comp4D}, one should consider the summation of the quantity ${\cal I}$ [defined in Eq. \eqref{calI}] over all possible values for the quantum number $m$. Because of the independence of the coefficient $g_{p,m}$ with respect to $m$, the introduction of an arbitrary weight function $f(m)$ satisfying constraint \eqref{constraint} is required to ensure the convergence of the sum over the integers $m$. This property illustrates somehow a redundancy character of the negative circulation quantum number, which is yet required to describe all possible phase singularities of the bicomplex geometry involving two circular angles.


\begin{thebibliography}{99}


\bibitem{Prange}
R. E. Prange and S. M. Girvin, {\em The Quantum Hall Effect} (Springer, New York, 1987).


\bibitem{Klitzing1980}
K. von Klitzing, G. Dorda, and M. Pepper, Phys. Rev. Lett. \textbf{45}, 494 (1980).


\bibitem{Tsui1982}
D. C. Tsui, H. L. Stormer, and A. C. Gossard, Phys. Rev. Lett. \textbf{48}, 1559 (1982).

\bibitem{Raikh1995}
M. E. Raikh and T. V. Shahbazyan, Phys. Rev. B \textbf{51}, 9682 (1995).

\bibitem{Champel2007}
T. Champel and S. Florens, Phys. Rev. B \textbf{75}, 245326 (2007).


\bibitem{Champel2008}
T. Champel, S. Florens, and L. Canet, Phys. Rev. B \textbf{78}, 125302 (2008).

\bibitem{Champel2009}
T. Champel and S. Florens, Phys. Rev. B \textbf{80}, 125322 (2009).

\bibitem{Champel2010}
T. Champel and S. Florens, Phys. Rev. B \textbf{82}, 045421 (2010).

\bibitem{Madelung1926}
E. Madelung, Z. Phys. \textbf{40}, 322 (1926).

\bibitem{Taka1983}
T. Takabayasi, Prog. Theor. Phys. \textbf{69}, 1323 (1983).

\bibitem{Malkin1969}
I. A. Malkin and V. I. Man'ko, Sov. Phys. JETP \textbf{28}, 527 (1969).

\bibitem{Glauber1963}
R. J. Glauber, Phys. Rev. \textbf{131}, 2766 (1963).

\bibitem{Zhang1990}
W.-M. Zhang, D. H. Feng, and R. Gilmore, Rev. Mod. Phys. \textbf{62}, 867 (1990).

\bibitem{Girvin}
S. M. Girvin and T. Jach, Phys. Rev. B \textbf{29}, 5617 (1984).

\bibitem{Kivelson}
S. Kivelson, C. Kallin, D. P. Arovas, and J. R. Schrieffer, Phys. Rev. B \textbf{36}, 1620 (1987).

\bibitem{Jain1987}
J. K. Jain and S. Kivelson, Phys. Rev. A \textbf{36}, 3467 (1987).

\bibitem{Jain1988}
J. K. Jain and S. Kivelson, Phys. Rev. B \textbf{37}, 4111 (1988).

\bibitem{Mermin}
N. D. Mermin, Rev. Mod. Phys. \textbf{51}, 591 (1979).


\bibitem{Sudarshan}
E. C. G. Sudarshan, Phys. Rev. Lett. \textbf{10}, 277 (1963).

%\bibitem{Roenn2001} S. R\"{o}nn, arXiv:math/0101200 (2001).

%\bibitem{Colombo2011} F. Colombo {\em et al.}, Ark. Mat. \textbf{49}, 277 (2011).

%\bibitem{DeBie2012} H. De Bie, D. C. Struppa, A. Vajiac, and M. B. Vajiac, Math. Nachr. \textbf{285}, 1230 (2012).

%\bibitem{Lavoie2010} R. G. Lavoie, L. Marchildon, and D. Rochon, Ann. Funct. Anal. \textbf{1}, 75 (2010).

%\bibitem{Lavoie2010b} R. G. Lavoie, L. Marchildon, and D. Rochon, AIP Conf. Proc. \textbf{1327}, 148 (2010).

% \bibitem{Lavoie2011} R. G. Lavoie, L. Marchildon, and D. Rochon, Adv. Appl. Clifford Algebras \textbf{21}, 561 (2011).

\bibitem{Davenport}
C. M. Davenport, {\em A Commutative Hypercomplex Calculus with Applications to Special Relativity} (Privately published, Knoxville, Tennessee, 1991).


\bibitem{Catoni2008}
F. Catoni, D. Boccaletti, R. Cannata, V. Catoni, E. Nichelatti, and P. Zampetti, {\em The Mathematics of Minkowski Space-Time: With an Introduction to Commutative Hypercomplex Numbers} (Birkh\"{a}user Verlag, Basel, 2008).

%\bibitem{Rochon2004} D. Rochon and S. Tremblay, Adv. Appl. Clifford Algebras \textbf{14}, 231 (2004).


%\bibitem{Rochon2006} D. Rochon and S. Tremblay, Adv. Appl. Clifford Algebras \textbf{16}, 135 (2006).

%\bibitem{Price} G. B. Price, {\em An Introduction to Multicomplex Spaces and Functions} (Marcel Dekker, New-York, 1990).

%\bibitem{Catoni2012} F. Catoni and P. Zampetti, Adv. Appl. Clifford Algebras \textbf{22}, 23 (2012).

%\bibitem{Stamp2012} P. C. E. Stamp, Phil. Trans. Roy. Soc. A \textbf{370}, 4429 (2012).

%\bibitem{Bassi2013} A. Bassi, K. Lochan, S. Satin, T. P. Singh, and H. Ulbricht, Rev. Mod. Phys. \textbf{85}, 471 (2013).

%\bibitem{Schloss} M. Schlosshauer, Rev. Mod. Phys. \textbf{76}, 1267 (2004).

\bibitem{zerodivisor} In abstract algebra, a zero divisor, $r$, of a ring is an element of the ring such that when multiplied by another non-zero element of the same ring satisfies $rs = sr=0$.

\bibitem{Philbin}
T. G. Philbin, Am. J. Phys. \textbf{82}, 742 (2014).


\bibitem{Adler1995}
S. Adler, {\em Quaternionic Quantum Mechanics and Quantum Fields} (Oxford University Press, New York, 1995).


\bibitem{Girvin1983}
S.M. Girvin and T. Jach, Phys. Rev. B \textbf{28}, 4506 (1983).


\bibitem{Girvin2004}
S. M. Girvin, {\em Introduction to the Fractional Quantum Hall Effect}, S\'{e}minaire Poincar\'{e}, 53 (2004).

\bibitem{Laughlin1983}
R. B. Laughlin, Phys. Rev. Lett. \textbf{50}, 1395 (1983).


\bibitem{Jain1989}
J. K. Jain, Phys. Rev. Lett. \textbf{63}, 199 (1989).


\bibitem{Jain2000}
J. K. Jain, Phys. Today \textbf{53} (4), 39 (2000).

\bibitem{Jain2007}
J. K. Jain, {\em Composite Fermions} (Cambridge University Press, Cambridge, 2007).

%\bibitem{Fogler1997} M. M. Fogler, A. Yu. Dobin, V. I. Perel, and B. I. Shklovskii, Phys. Rev. B \textbf{56}, 6823 (1997).


%\bibitem{Hashi2008} K. Hashimoto, C. Sohrmann, J. Wiebe, T. Inaoka, F. Meier, Y. Hirayama, R. A. R\"{o}mer, R. Wiesendanger, and M. Morgenstern, Phys. Rev. Lett. \textbf{101}, 256802 (2008).


%\bibitem{Hashi2012} K. Hashimoto, T. Champel, S. Florens, C. Sohrmann, J. Wiebe, Y. Hirayama, R. A. R\"{o}mer, R. Wiesendanger, and M. Morgenstern, Phys. Rev. Lett. \textbf{109}, 116805 (2012).


%\bibitem{Floser2013} M. Fl\"{o}ser, B. A. Piot, C. L. Campbell, D. K. Maude, M. Henini, R. Airey, Z. R. Wasilewski, S. Florens, and T. Champel, New J. Physics. \textbf{15}, 083027 (2013).

\bibitem{Ruiz}
S.  M. Ruiz, The Mathematical Gazette {\bf 80}, 579 (1996).

\end{thebibliography}
\end{document}